\documentclass{appolb}
\usepackage{graphicx}
\usepackage{amsmath}

\begin{document}
\title{Finite temperature Mott transition in a nonlocal PNJL model%
\thanks{Presented at Excited QCD 2013}%
}
\author{Sanjin Beni\' c
\address{Physics Department, Faculty of Science, 
University of Zagreb, Zagreb, Croatia}
\and
David Blaschke
\address{Institut Fizyki Teoretycznej, 
Uniwersytet Wroc{\l}awski, Wroc{\l}aw, Poland}
\address{Bogoliubov Laboratory for 
Theoretical Physics, JINR Dubna, Dubna, Russia}
\address{Fakult\"at f\"ur Physik, Universit\"at Bielefeld,
Bielefeld, Germany}
}
\maketitle
\begin{abstract}
We provide a novel calculation of the Mott effect in
non-local PNJL models.
We find that the ``deconfinement" transition temperature in these models 
is lower than the Mott temperature.
Furthermore, the mass and the width of the $\sigma$ and the $\pi$ meson 
modes is calculated with the result that the width in nonlocal models is 
in general reduced as compared to local models.
Difficulties encountered while attempting to ``Wick rotate'' covariant models
are carefully discussed.
\end{abstract}
  
\section{Introduction}

\begin{center}
\begin{figure}[htb]
\centerline{%
\includegraphics[width=10.0cm]{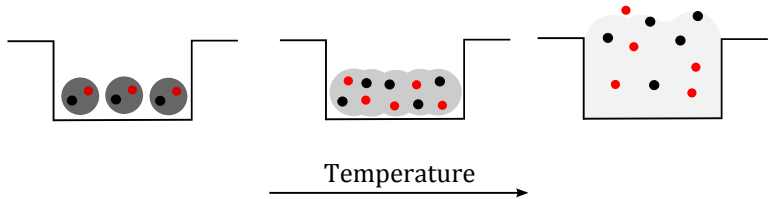}}
\caption{(Color online) An illustration of the Mott transition.
The bound states represented by shaded area are shown in a schematic potential
well.
Color brigthness accounts for the strength of the interaction.}
\label{fig:mott}
\end{figure}
\end{center}

It is by now established that the first rise in the QCD 
pressure as calculated
from the lattice is well described by the 
Hadron Resonance Gas 
\cite{Karsch:2003zq,Borsanyi:2010bp}.
This important result offers a simple physical picture in understanding
a phase transition from the hadronic world to the quark-gluon plasma:
with an increase of the temperature, the meson wave 
functions start to overlap.
Due to Pauli blocking, quarks are then 
forced to occupy higher quantum ``states" finally to merge
into the continuum, see Fig. \ref{fig:mott}.
More over, it calls for a description of the lattice data within
a unified field-theoretical model where the meson degrees of freedom
are interpreted as true $\bar{q}q$ bound states.
Such a microscopic description might be vital 
for understanding the quark-hadron
transition in general \cite{Blaschke:1984yj}.

In this contribution we provide a brief explorative account 
of this \textit{Mott transition} in a 
covariant, nonlocal, Polyakov-Nambu-Jona--Lasinio (nl-PNJL) 
model \cite{Horvatic:2010md,Radzhabov:2010dd,Hell:2011ic,Carlomagno:2013ona}.
Quark models in general have a potential to offer a powerful
theoretical setup for a systematic  introduction of fluctuations beyond the
mean field, leading to a virial expansion. 
Important progress in that direction has been achieved recently 
\cite{Radzhabov:2010dd,Wergieluk:2012gd,Yamazaki:2012ux,Blaschke:2013zaa}.

\section{Selected features of the model}

\begin{center}
\begin{figure}[htb]
\centerline{%
\includegraphics[width=7.0cm]{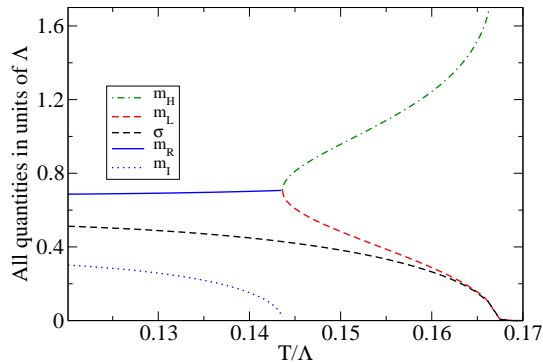}}
\caption{(Color online) Temperature dependence of the lowest lying 
CCMPs in the chiral limit and without the Polyakov loop.
The imaginary part of the singularity vanishes when the temperature rises to 
the point where $\sigma(T_\mathrm{cont})=\sigma_c = \Lambda/\sqrt{2e}$.}
\label{fig:sing}
\end{figure}
\end{center}

The nl-PNJL models are characterized by  
the running of the mass function in the quark propagator
provided by $M(p^2)=m+\sigma f(p^2)$.
In order to properly address the Mott transition it 
is important to study the analytic properties of 
the quark propagator.
The latter is highly non-trivial due to 
the fact that the mass ``runs'' in a covariant fashion.
Utilizing a Gaussian form factor $f(p^2) = e^{-p^2/\Lambda^2}$ 
it is easy to show that the propagator exhibits an infinite sequence 
of complex conjugate mass poles (CCMPs),
if the gap $\sigma$ is above the critical value $\sigma_c=\Lambda/\sqrt{2e}$.
For $\sigma<\sigma_c$, the complex pole, together with its 
complex conjugate counterpart ``fuses" into a doublet of 
real poles $zz^*\to HL$, as shown
in Fig. \ref{fig:sing}.
In the chiral limit ($m=0$) real poles are provided by the Lambert $W$-function
\begin{equation}
m_{L,H}^2(\sigma) = -\frac{1}{2} W_{0,-1}(-2\sigma^2/\Lambda^2)~,
\end{equation}
where the ``heavy" state is non-physical in the sense $m_H\to\infty$.
By contrast, the ``light" state quickly joins the 
mass gap $\sigma$ at high $T$ where it can be interpreted 
as a physical state, see Fig.~\ref{fig:sing}.

For the model parameters 
which we employ here \cite{GomezDumm:2006vz}
always holds $\sigma>\sigma_c$. 
Therefore, the physical continuum of states 
appears only at temperatures above
$T>T_\mathrm{cont}$.

\section{Finite temperature $\pi$ and $\sigma$ correlations}
The central object of our study is the meson
polarization function 
\cite{Blaschke:2000gd,Scarpettini:2003fj,Horvatic:2007wu,Contrera:2009hk} 
\begin{equation}
\Pi_M (\nu_m,|\mathbf{q}|)= \frac{2d_q}{3} T\sum_n \int\frac{d^3 p}{(2\pi)^3}
\mathrm{tr}_C\left[f^2(p_n^2)
\frac{K_M(\omega_n^2,\mathbf{p}^2,\nu_m^2,\mathbf{q}^2)}
{\mathcal{D}((p_n^-)^2)
\mathcal{D}((p_n^+)^2}\right]~,
\label{pol}
\end{equation}
with $d_q = 2\times 2 \times N_f\times N_c$ being the quark degrees of freedom, and
\begin{equation}
K_M(\omega_n^2,\mathbf{p}^2,\nu_m^2,\mathbf{q}^2)=
(p_n^+ p_n^-)\pm
M((p_n^+)^2)M((p_n^-)^2)~.
\label{gem}
\end{equation}
We use the following notation: $M=\pi,\sigma$, 
$q_m =(\nu_m,\mathbf{q})$, where
$\nu_m = 2m\pi T$ are the bosonic Matsubara frequencies.
Furthermore, $p_n^{\pm} = (\omega_n^{\pm},\mathbf{p}^\pm)$,
with $\omega_n^{\pm} = \omega_n \pm \nu_m/2$, and
$\mathbf{p}^\pm = \mathbf{p}\pm \mathbf{q}/2$ and 
\begin{equation}
\mathcal{D}(-z^2,\mathbf{p}^2)=
-z^2+\mathbf{p}^2 +
M^2(-z^2+\mathbf{p}^2)~.
\label{denom}
\end{equation}

\subsection{Imaginary part of the polarization function}

In a first attempt to discuss the Mott effect, we consider the meson masses 
as given by the spatial, or screening masses, from the respective 
Bethe-Salpeter equations
$1-G_S\Pi_M(\nu_m =0,|\mathbf{q}|=-i m_M) = 0$.
while for the widths we calculate
the imaginary part of (\ref{pol}).
Details of this calculation are provided in \cite{Benic:2013tba}.
For the meson at rest $\mathbf{q}=0$ we find
\begin{equation}
\begin{split}
\mathrm{Im}[\Pi_M(-iq_0,0)] &= \frac{d_q}{16\pi}
\left[1- 2 n_\Phi(q_0/2)\right]
\sqrt{1-\left(\frac{2m_L}{q_0}\right)^2}f^2\left(\frac{q_0^2}{4}-m_L^2\right)\\
&\times 
\frac{K_M\left(0,\frac{q_0^2}{4}-m_L^2,-q_0^2,0\right)}
{\left[\mathcal{D}'
\left(-\frac{q_0^2}{4},\frac{q_0^2}{4}-m_L^2\right)\right]^2}
~\theta\left(\frac{q_0}{2}-m_L\right)~,
\end{split}
\label{img_pol}
\end{equation}
where $n_\Phi(z)$ is the generalized occupation number function 
in the presence of the Polyakov loop $\Phi$.
For details see, e. g., Ref.~\cite{Hansen:2006ee}.
In the local limit $f\to 1$, it is easy to show that the standard PNJL result 
\cite{Hansen:2006ee}is correctly reproduced .
In deriving this formula we have ignored the threshold for the meson to decay 
to a heavy state $M\to HH$, as well as possible mixed channels $H\to M L$.
For the meson width we use
\begin{equation}
\Gamma_M = g_{M\bar{q}q}\frac{\mathrm{Im}[\Pi_M]}{m_M} ~,
\label{wid}
\end{equation}
where $g_{M\bar{q}q}$ is the residue of the meson propagator.
In numerical calculations we assume that the 
leading effect of $g_{M\bar{q}q}$
is to cancel the $f^2$ term in $\mathrm{Im}[\Pi_M]$.

\section{Results and conclusion}
\begin{center}
\begin{figure}[htb]
\centerline{%
\includegraphics[width=8.0cm,height=5cm]{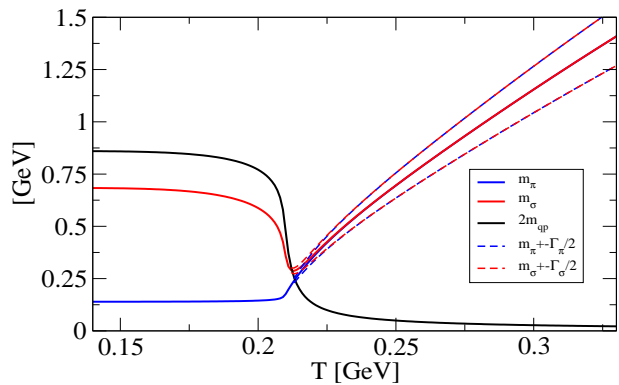}}
\caption{(Color online)  Masses of $\pi$ (blue) and $\sigma$ meson (red).
The ``continuum" threshold $2m_\mathrm{qp}$ is shown in black.
The range of a half-width around the meson mass is given by dashed lines.}
\label{fig:fluc}
\end{figure}
\end{center}

In Fig.~\ref{fig:fluc} we present results of our approach.
Besides the spatial meson masses and widths, the ``continuum'' 
states defined by $2m_\mathrm{qp}$, where $m_\mathrm{qp}=m+\sigma$, 
are also shown.
Strictly speaking, these states are not present as 
real singularities of the quark
propagator up to $T_\mathrm{cont}$, giving a 
distinctive feature of covariant models.
Nevertheless, at $T>T_\mathrm{cont}$, they are almost identical to the
actual singularities, see Fig. \ref{fig:sing}, and therefore provide an 
intuitive picture also in the low temperature region.

As the temperature increases the continuum threshold drops, eventually
hitting the meson masses at the Mott temperature $T_\mathrm{Mott}$.
From that point on the sharp meson states become resonances by 
acquiring a finite width.
Around the same temperature the $\sigma$ and the $\pi$ meson become 
degenerate thus making the restoration of the chiral symmetry manifest.

In contrast to the local PNJL calculation, see e. g. \cite{Hansen:2006ee},
we find a smaller width, and a significantly higher mass, 
their ratio being even around $\Gamma_M/M_M \sim 0.1$
at $T\sim 300$ MeV, see Fig \ref{fig:fluc}.
The latter is a common feature of spatial masses.

While the presented calculation already reveals some 
aspects of the nonlocality
of the interaction, namely a significant reduction of the width of the resonance, 
the crucial step in properly accounting for the Mott
transition in covariant models is still lacking.
While at low $T$ we expect bound states to dominate the thermodynamics,
the 2-particle correlation contribution in the high temperature regime
shall come from the quark-antiquark scattering \cite{Wergieluk:2012gd,Yamazaki:2012ux,Blaschke:2013zaa}.

We find that in covariant models such an approach is hindered by the additional
(unphysical) singularities, making the original physical picture blurry.
For example to account for the scattering, one needs to put quarks on shell 
(which is possible after $T_\mathrm{cont}$).
This calls for a Wick rotation of the effective interaction itself.
At least with a Gaussian regulator such an approach would yield a term 
$e^{q_0^2/\Lambda^2}$, where $q_0$ is the typical energy of the process.
Thus, when $q_0 \sim T \sim \Lambda$, scattering would seemingly grow without 
bound and eventually violate unitarity.

It is then clear that knowing the effective dressing of the quarks only in 
Euclidean space is insufficient.
Moreover, one easily imagines that by using different analytic choices for the 
regulators in Euclidean space,
these might behave quite differently in the complex energy plane, although being 
qualitatively same in Euclidean space.
We believe that a more appropriate ``gauge'' for fully accounting the physics
of the Mott transition might be the Coulomb gauge 
\cite{Guo:2009ma,Pak:2011wu,Watson:2012ht}, or its covariant formulations 
\cite{Morozov:2002hv}, where the interaction does not depend on the energy.

\end{document}